\documentclass[12pt]{article}

\usepackage{cite}
\usepackage{a4}

\newcommand{\ol}{\overline}
\newcommand{\ul}{\underline}
\newcommand{\tr}{{\rm tr}}
\newcommand{\wt}{\widetilde}
\newcommand{\wh}{\widehat}

\begin{document}


\begin{titlepage}
\title{\hfill\parbox{4cm}
       {\normalsize UT-08-23\\July 2008}\\
       \vspace{1.5cm}
${\cal N}=4$ Chern-Simons theories\\with auxiliary vector multiplets
       \vspace{1.5cm}}
\author{
Yosuke Imamura\thanks{E-mail: \tt imamura@hep-th.phys.s.u-tokyo.ac.jp}\quad
and\quad
Keisuke Kimura\thanks{E-mail: \tt kimura@hep-th.phys.s.u-tokyo.ac.jp}
\\[20pt]
{\it Department of Physics, University of Tokyo,}\\ {\it Tokyo 113-0033, Japan}
}
\date{}

\maketitle
\thispagestyle{empty}

\vspace{0cm}

\begin{abstract}
\normalsize
We investigate a class of quiver-type Chern-Simons gauge theories
with some Chern-Simons couplings vanishing.
The vanishing of the couplings means
that the corresponding vector fields
are auxiliary fields.
We show that these theories possess ${\cal N}=4$ supersymmetry
by writing down the actions
and the supersymmetry transformation
in terms of component fields in
manifestly $Spin(4)$ covariant form.
\end{abstract}

\end{titlepage}

\clearpage
\section{Introduction}
Recently,
Bagger, Lambert\cite{Bagger:2006sk,Bagger:2007jr,Bagger:2007vi},
and Gusstavson\cite{Gustavsson:2007vu,Gustavsson:2008dy}
proposed a new field theory model
as a promising candidate
for the theory describing multiple M2-branes.
This model (BLG model) is based on Lie $3$-algebras,
and can also be regarded as a special class of
Chern-Simons gauge theories\cite{Bandres:2008vf,VanRaamsdonk:2008ft}
with ${\cal N}=8$ supersymmetry.

Until quite recent, the largest known supersymmetry of interacting Chern-Simons theories
had been ${\cal N}=3$.
This is because supersymmetric completion of
Chern-Simons terms include bi-linear terms of superpartners of gauge fields
which break R-symmetry down to $SO(3)$
(or $Spin(3)$ when hyper multiplets are present).
See \cite{Gaiotto:2007qi} for detailed analysis of ${\cal N}=2,3$ superconformal Chern-Simons theories.

If the Yang-Mills kinetic term is absent, the situation changes.
In such a case superpartners of gauge fields become
non-dynamical auxiliary fields,
and there is a possibility that the R-symmetry enhances
when these auxiliary fields are integrated out.
The ${\cal N}=8$ supersymmetry of
the BLG model is a special case of such symmetry enhancement.
The BLG model is very restricted,
and if we require the algebra is finite dimensional and
has positive definite metric,
the only possible gauge group is $SO(4)$\cite{Papadopoulos:2008sk,Gauntlett:2008uf}.
(The positivity of the metric is not indispensable for
the consistency of the theory.
See \cite{Gomis:2008uv,Benvenuti:2008bt,Ho:2008ei,Bandres:2008kj,Gomis:2008be}.)

In the case of ${\cal N}<8$, we have larger variety of
theories.
Gaiotto and Witten\cite{Gaiotto:2008sd} showed that
the supersymmetry can be enhanced to ${\cal N}=4$
in a class of Chern-Simons theories with product gauge groups
$U(N)\times U(N')$ and $Sp(N)\times SO(N')$.
This is generalized in \cite{Hosomichi:2008jd} to quiver type gauge theories
by introducing twisted hypermultiplets.
They construct ${\cal N}=4$ Chern-Simons theories described by linear and circular quiver diagrams.
A $U(N)\times U(N)$ Chern-Simons theory with ${\cal N}=6$ supersymmetry
is also proposed
in \cite{Aharony:2008ug}.
For recent progress in ${\cal N}\geq4$ Chern-Simons theories,
see also
\cite{Fuji:2008yj,Benna:2008zy,Bhattacharya:2008bj,Nishioka:2008gz,Honma:2008jd,Imamura:2008nn,Minahan:2008hf,Hanany:2008qc,Gaiotto:2008cg,Grignani:2008is,Hosomichi:2008jb,Bagger:2008se,Terashima:2008sy,Grignani:2008te,Terashima:2008ba,Gromov:2008bz,Ahn:2008gd,Gromov:2008qe,Chen:2008qq,Cherkis:2008qr,Bandres:2008ry,Gomis:2008vc,Schnabl:2008wj,Li:2008ya,Kim:2008gn,Pang:2008hw,Garousi:2008ik,Hashimoto:2008iv,Astolfi:2008ji}.

In this paper we investugate a class of ${\cal N}=4$ Chern-Simons theories.
The model is described by a circular quiver diagram with circumference $n$.
Namely, gauge group is $\prod_{I=1}^nU(N_I)$, and there are $n$ hypermultiplets
belonging to bi-fundamental representations.
The action of this model is
\begin{equation}
S=S_{\rm CS}+S_{\rm hyper},
\label{model}
\end{equation}
where $S_{\rm CS}$ and $S_{\rm hyper}$ are given in terms of ${\cal N}=2$ superfields
by
\begin{equation}
S_{\rm CS}
=\sum_{I=1}^n
k_I\tr\left[\int d^3xd^4\theta
\left(-\frac{i}{2}D V_I\ol D V_I+\cdots\right)
+\left(-\frac{i}{2}\int d^3xd^2\theta\Phi_I^2+\mbox{c.c.}\right)
\right],
\label{ScsSF}
\end{equation}
and
\begin{eqnarray}
S_{\rm hyper}
&=&-\sum_{I=1}^n\int d^3xd^4\theta\tr(\ol Q_Ie^{2V_I}Q_Ie^{-2V_{I+1}}+\wt Q_Ie^{-2V_I}\ol{\wt Q}_Ie^{2V_{I+1}})
\nonumber\\&&
+\sum_{I=1}^n\left(\int d^3xd^2\theta\sqrt2i
\tr(\wt Q_I\Phi_I Q_I- Q_I\wt Q_I\Phi_{I+1})
+\mbox{c.c.}\right)
\label{ShyperSF}.
\end{eqnarray}
A brief summary of ${\cal N}=2$ superfield formalism
is given in Appendix \ref{sf.sec}.
The $n$ vector and $n$ hyper
multiplets are labeled by the same index $I$.
$I=n+1$ is identified with $I=1$.
$V_I$ and $\Phi_I$ are an ${\cal N}=2$ vector and an adjoint chiral superfield,
respectively,
and they form an ${\cal N}=4$ vector multiplet.
$Q_I$ and $\wt Q_I$ are bi-fundamental chiral superfields
belonging to $({\bf N}_I,\ol {\bf N}_{I+1})$ and $(\ol {\bf N}_I,{\bf N}_{I+1})$ of $U(N_I)\times U(N_{I+1})$,
and these form an ${\cal N}=4$ hypermultiplet.

If the Chern-Simons coupling $k_I$ of $U(N_I)$ is $k_I=(-)^Ik$,
this theory coincides with a model proposed in \cite{Hosomichi:2008jd}.
We extend the model by considering
more general Chern-Simons couplings
\begin{equation}
k_I=\frac{k}{2}(s_I-s_{I-1}),\quad
s_I=\pm1,\quad
k>0.
\label{ki}
\end{equation}
The model in \cite{Hosomichi:2008jd} corresponds to
the choice $s_I=(-1)^I$.
We allow $s_I$ to be $\pm1$ in arbitrary order.
This implies that we allow some of Chern-Simons couplings to vanish.
If $k_I=0$, all the component fields of $V_I$ and $\Phi_I$ become auxiliary fields.
We call such multiplets ``auxiliary vector multiplets.''
For distinction we call
vector multiplets with $k_I\neq0$
``dynamical vector multiplets''
although they have no propagating degrees of
freedom.

Chern-Simons theories
with such auxiliary vector multiplets
are discussed by Gaiotto and Witten in \cite{Gaiotto:2008sd}.
They introduce such multiplets to define non-trivial
hyper-K\"ahler manifolds
as hyper-K\"ahler quotients.
By integrating out the auxiliary vector multiplets
in our model we obtain
a Chern-Simons gauge theory
coupling to sigma models
with hyper-K\"ahler target spaces.
This model is similar to the model in \cite{Hosomichi:2008jd},
but hyper and twisted hyper multiplets in the model
are replaced by non-trivial sigma models.

The purpose of this paper is to show
that our model
possesses $Spin(4)$ R-symmetry and ${\cal N}=4$ supersymmetry.
It would be possible to prove it
by extending the arguments in \cite{Hosomichi:2008jd}
by generalizing minimally coupled matter fields
to general hyper-K\"ahler sigma models.
In this paper, however, we adopt different way of proof.
We integrate out only the auxiliary fields in the hyper and dynamical vector multiplets,
and leave the component fields in the auxiliary vector multiplets in the action.
A good point of this treatment is that we do not have to
solve the non-linear constraints imposed on the moment maps
for auxiliary gauge fields.
We will show in the following sections that,
after integrating out the auxiliary fields
in hyper and dynamical vector multiplets,
the action (\ref{model}) can be
rewritten in manifestly $Spin(4)$ invariant form.
Because ${\cal N}=2$ supersymmetry of our model
is manifest by construction,
the $Spin(4)$ invariance of the action
implies that the existence of
${\cal N}=4$ supersymmetry.

The expression of Chern-Simons couplings $k_I$ in (\ref{ki}) is closely related
to a brane construction of the model.
Our model is the low energy limit of the theory realized
on a brane system in type IIB string theory.
It consists of a stack of $N$ D3-branes wrapped on ${\bf S}^1$
and $n$ fivebranes
intersecting with the D3-branes.
We label the fivebranes by $I=1,\ldots,n$
in order of intersections with the D3-branes
along ${\bf S}^1$.
If the charge of $I$-th fivebrane is $(m_I,1)$,
the Chern-Simons coupling
of the gauge field living on the interval
of the D3-branes between
two intersections $I$ and $I-1$ is given by\cite{Kitao:1998mf,Bergman:1999na}
\begin{equation}
k_I=\frac{1}{2\pi}(m_I-m_{I-1}).
\end{equation}
If there are only two types of fivebranes,
the Chern-Simons couplings are given by (\ref{ki}).

The action of gauge theory realized on this brane system
is $S_{\rm YM}+S_{\rm CS}+S_{\rm hyper}$
where $S_{\rm CS}$ and $S_{\rm hyper}$ are given in
(\ref{ScsSF}) and (\ref{ShyperSF}), respectively,
and $S_{\rm YM}$
includes
the Yang-Mills kinetic terms.
It is given by
\begin{equation}
S_{\rm YM}
=\sum_{I=1}^n\frac{1}{g_I^2}
\left[\frac{1}{2}\int d^3x d^2\theta\tr W_I^2
-\int d^3xd^4\theta \tr(\ol\Phi_Ie^{2V_I}\Phi e^{-2V_{I+1}})
\right],
\label{SymSF}
\end{equation}
where $g_I$ is Yang-Mills gauge couplings
depending on the position of intersecting points of branes.
The brane system preserves ${\cal N}=3$ supersymmetry,
which coincides with the supersymmetry of the
Yang-Mills-Chern-Simons action
$S_{\rm YM}+S_{\rm CS}+S_{\rm hyper}$.

In the low energy limit, the kinetic terms in $S_{\rm YM}$ become irrelevant
because the coupling constants $g_I$ have mass dimension $1/2$.
The supersymmetry enhancement in this limit
is strongly suggested by an analysis of moduli space.
The Higgs branch of this model is studied in
\cite{Imamura:2008nn}, and
it is shown that the moduli space for $N_I=1$ is
an orbifold in the form
\begin{equation}
{\bf C}^4/\Gamma,
\end{equation}
where $\Gamma$ is a certain discrete subgroup
consisting of elements of the form
\begin{equation}
(z_1,z_2,z_3,z_4)\rightarrow
(e^{i\alpha}z_1,e^{-i\alpha}z_2,e^{i\beta}z_3,e^{-i\beta}z_4).
\end{equation}
If we assume the flat metric,
this orbifold preserves ${\cal N}=4$ supersymmetry.

This paper is organized as follows.
In the next section we rewrite the actions
given above in terms of component fields.
It makes $Spin(4)$ R-symmetry and ${\cal N}=4$ supersymmetry
of Yang-Mills-matter system $S_{\rm YM}+S_{\rm hyper}$ manifest.
We emphasize that these symmetries are different from
those of the Chern-Simons-matter system $S_{\rm CS}+S_{\rm hyper}$.
In order to distinguish the symmetries of these two
systems, we denote the $Spin(4)$ R-symmetry and ${\cal N}=4$ supersymmetry
of the Yang-Mills-matter system
by $R_{\rm YM}$ and ${\cal N}=4_{\rm YM}$,
while we refer to those of Chern-Simons theory
as $R_{\rm CS}$ and ${\cal N}=4_{\rm CS}$.
In \S\ref{susy.sec}
${\cal N}=4_{\rm CS}$ supersymmetry transformation is written down
in manifestly $R_{\rm CS}$ covariant form.
In \S\ref{action.sec} we prove the $R_{\rm CS}$ invariance
of the action $S_{\rm CS}+S_{\rm hyper}$.
\S\ref{conc.sec} is the concluding section.

\section{Action in terms of component fields}
In this section we rewrite the actions
given in the introduction in terms of component fields.
This makes $R_{\rm YM}=Spin(4)$ R-symmetry of $S_{\rm YM}$ and $S_{\rm hyper}$
and $Spin(3)$ R-symmetry of $S_{\rm CS}$ manifest.

Let us first rewrite the Yang-Mills action $S_{\rm YM}$
in (\ref{SymSF}).
Although this vanishes
in the low-energy limit $g_I\rightarrow\infty$
and irrelevant to our model,
it may be instructive to know
the explicit form of this action.
It is given by
\begin{eqnarray}
S_{\rm YM}
&=&
\sum_{I=1}^n\frac{1}{g_I^2}
\int d^3x\tr\left[
-\frac{1}{4}F_{I\mu\nu}F_I^{\mu\nu}
+\frac{i}{2}\lambda_I^{A\dot B}\gamma^\mu D_\mu\lambda_{IA\dot B}
-\frac{1}{4}D_\mu\phi_I^{\dot A}{}_{\dot B} D^\mu\phi_I^{\dot B}{}_{\dot A}
\right.
\nonumber\\&&
\left.
-\frac{i}{2}\lambda_{IA\dot B}[\phi_I^{\dot B}{}_{\dot C},\lambda_I^{A\dot C}]
+
\frac{1}{4}F_I^A{}_BF_I^B{}_A
+\frac{1}{16}[\phi_I^{\dot A}{}_{\dot B},\phi_I^{\dot C}{}_{\dot D}]
[\phi_I^{\dot B}{}_{\dot A},\phi_I^{\dot D}{}_{\dot C}]
\right].
\label{Sym}
\end{eqnarray}
This includes
$U(N_I)$ gauge fields $F_{I\mu\nu}$,
fermions $\lambda_I^{A\dot B}$,
scalars $\phi_I^{\dot A}{}_{\dot B}$,
and auxiliary fields $F_I^A{}_B$.
All these fields belong to the adjoint representation
of $U(N_I)$,
and satisfy the reality conditions
\begin{equation}
(F_{I\mu\nu})^\dagger
=F_{I\mu\nu},\quad
(\lambda_I^{A\dot B})^\dagger
=-\lambda_{IA\dot B},\quad
(\phi_I^{\dot A}{}_{\dot B})^\dagger
=\phi_I^{\dot B}{}_{\dot A},\quad
(F_I^A{}_B)^\dagger
=F_I^B{}_A.
\end{equation}
We raise and lower pairs of $SU(2)$ indices of bi-spinors by
the relation
\begin{equation}
\lambda_{IA\dot B}
=\epsilon_{AC}\epsilon_{\dot B\dot D}\lambda_I^{C\dot D},\quad
\epsilon_{12}=\epsilon^{12}=\epsilon_{\dot1\dot2}=\epsilon^{\dot1\dot2}=1.
\end{equation}
$\phi_I$ and $F_I$ are traceless
\begin{equation}
\phi_I^{\dot A}{}_{\dot A}=F_I^A{}_A=0.
\end{equation}
This action possesses global
$R_{\rm YM}=SU(2)_L\times SU(2)_R$ symmetry.
$SU(2)_L$ and $SU(2)_R$ act on
undotted indices $A,B,\ldots=1,2$
and dotted ones $\dot A,\dot B,\ldots=\dot1,\dot2$,
respectively.

The action of hypermultiplets $S_{\rm hyper}$ in
(\ref{ShyperSF}) is rewritten as
\begin{eqnarray}
S_{\rm hyper}
&=&
\sum_{I=1}^n
\int d^3x\tr\Big[
-D_\mu\ol q_{IA} D^\mu q_I^A
-i\ol\psi_I^{\dot A}\gamma^\mu D_\mu\psi_{I\dot A}
-F_I^A{}_B(\mu_I^B{}_A-\wt\mu_{I-1}^B{}_A)
\nonumber\\
&&
-i\lambda_{IA\dot B}(j_I^{A\dot B}-\wt j_{I-1}^{A\dot B})
+i\psi_{I\dot B}\ol\psi_I^{\dot A}\phi_I^{\dot B}{}_{\dot A}
-i\ol\psi_{I-1}^{\dot A}\psi_{I-1\dot B}\phi_I^{\dot B}{}_{\dot A}
\nonumber\\&&
-\frac{1}{2}\nu_I^A{}_A\phi_I^{\dot B}{}_{\dot C}\phi_I^{\dot C}{}_{\dot B}
-\frac{1}{2}\wt\nu_{I-1}^A{}_A\phi_I^{\dot B}{}_{\dot C}\phi_I^{\dot C}{}_{\dot B}
+\ol q_{IA}\phi_I^{\dot B}{}_{\dot C}q_I^A\phi_{I+1}^{\dot C}{}_{\dot B}
\Big].
\label{Shyper}
\end{eqnarray}
This includes scalar fields $q_I$ and fermions $\psi_I$.
The auxiliary fields in $Q_I$ and $\wt Q_I$ were integrated out
so that the $R_{\rm YM}$ symmetry becomes manifest.
We defined bi-linears
\begin{equation}
\nu_I^A{}_B=q_I^A\ol q_{IB},\quad
\wt\nu_I^A{}_B=\ol q_{IB}  q_I^A,
\end{equation}
\begin{equation}
\mu_I^A{}_B
=\nu_I^A{}_B-\tr
=\nu_I^A{}_B-\frac{1}{2}\nu_I^C{}_C\delta^A_B,\quad
\wt\mu_I^A{}_B
=\wt\nu_I^A{}_B-\tr
=\wt\nu_I^A{}_B-\frac{1}{2}\wt\nu_I^C{}_C\delta^A_B,
\label{mudef}
\end{equation}
and
\begin{equation}
j_I^{A\dot B}=\sqrt2q_I^A\ol\psi_I^{\dot B}
-\sqrt2\epsilon^{AC}\epsilon^{\dot B\dot D}\psi_{I\dot D}\ol q_{IC},
\quad
\wt j_I^{A\dot B}=\sqrt2\ol\psi_I^{\dot B}q_I^A
-\sqrt2\epsilon^{AC}\epsilon^{\dot B\dot D}\ol q_{IC}\psi_{I\dot D}.
\label{jdef}
\end{equation}
``$-\tr$'' used in (\ref{mudef})
represents
the subtraction of the trace part of two $SU(2)$ indices.
(\ref{mudef}) and (\ref{jdef})
are components of current multiplets
coupled by the vector multiplets.
Other components in the multiplets and
the supersymmetry transformation of the components
are given in Appendix \ref{current.sec}.
Indices in (\ref{Shyper}) are consistently
contracted, and this action is manifestly $R_{\rm YM}$ invariant.
The $R_{\rm YM}$ representations of component fields
are summarized in Table \ref{rep.tbl}.
\begin{table}[htb]
\caption{$R_{\rm YM}=SU(2)_L\times SU(2)_R$
representations of component fields
in the ${\cal N}=4$ supersymmetric
Yang-Mills-matter system
are shown.}
\label{rep.tbl}
\begin{center}
\begin{tabular}{cccc|cc}
\hline
\hline
$v_{I\mu}$ & $\phi_I$ & $\lambda_I$ & $F_I$ & $q_I$ & $\psi_I$ \\
\hline
$({\bf1},{\bf1})$ & $({\bf1},{\bf3})$ & $({\bf2},{\bf2})$ & $({\bf3},{\bf1})$ & $({\bf2},{\bf1})$ & $({\bf1},{\bf2})$ \\
\hline
\end{tabular}
\end{center}
\end{table}

The ${\cal N}=4_{\rm YM}$ supersymmetry transformation is given by
\begin{eqnarray}
\delta\phi_I^{\dot A}{}_{\dot B}
&=&
2i(\xi_{C\dot B}\lambda_I^{C\dot A})
-i\delta^{\dot A}_{\dot B}(\xi_{B\dot C}\lambda_I^{B\dot C}),\\
\delta v_{I\mu}
&=&-(\xi_{A\dot B}\gamma_\mu\lambda_I^{A\dot B}),
\label{deltav0}\\
\delta\lambda_I^{A\dot B}
&=&\frac{i}{2}\gamma^{\mu\nu}\xi^{A\dot B} F_{I\mu\nu}
+\gamma^\mu\xi^{A\dot C} D_\mu\phi_I^{\dot B}{}_{\dot C}
+F_I^A{}_C\xi^{C\dot B}
+\frac{1}{2}[\phi_I^{\dot B}{}_{\dot C},\phi_I^{\dot C}{}_{\dot D}]\xi^{A\dot D},\\
\delta F_I^A{}_B
&=&
2i(\xi_{B\dot C}\gamma^\mu D_\mu\lambda_I^{A\dot C})
-2i(\xi_{B\dot C}[\phi_I^{\dot C}{}_{\dot D},\lambda_I^{A\dot D}])
-\tr,
\end{eqnarray}
for vector multiplets and
\begin{eqnarray}
\delta q_I^A
&=&\sqrt2i(\xi^{A\dot B}\psi_{I\dot B}),
\label{deltaq}\\
\delta\psi_{I\dot A}
&=&
\sqrt2\xi_{C\dot B}\phi_I^{\dot B}{}_{\dot A} q_I^C
-\sqrt2\xi_{C\dot B}q_I^C\phi_{I+1}^{\dot B}{}_{\dot A} 
+\sqrt2\gamma^\mu\xi_{B\dot A} D_\mu q_I^B,
\label{deltapsi}
\end{eqnarray}
for hyper multiplets.
The parameter $\xi^{A\dot B}$ belongs to $({\bf 2},{\bf 2})$
representation of $R_{\rm YM}=SU(2)_L\times SU(2)_R$.

The introduction of Chern-Simons terms $S_{\rm CS}$ in (\ref{ScsSF})
breaks the supersymmetry to ${\cal N}=3$.
We can see this by rewriting the action in terms of component fields.
\begin{eqnarray}
S_{\rm CS}
&=&\sum_{I=1}^n
k_I\int d^3x\tr\left[
\epsilon^{\mu\nu\rho}
\left(\frac{1}{2}v_{I\mu}\partial_\nu v_{I\rho}-\frac{i}{3}v_{I\mu} v_{I\nu} v_{I\rho}\right)
\right.
\nonumber\\
&&\left.
+\frac{1}{2}\phi_I^{\dot A}{}_{\dot B}F_I^B{}_A
+\frac{1}{6}\phi_I^{\dot A}{}_{\dot B}\phi_I^{\dot B}{}_{\dot C}\phi_I^{\dot C}{}_{\dot A}
+\frac{i}{2}\lambda_I^{A\dot B}\lambda_{IB\dot A}
\right].
\label{Scs}
\end{eqnarray}
In this action,
some dotted indices are contracted with undotted indices,
and thus $R_{\rm YM}$
is broken to its diagonal subgroup $SU(2)_D$.
The parameter $\xi^{A\dot B}$ is split into
the singlet and the triplet of $SU(2)_D$,
and only the triplet part of the supersymmetry
is preserved by the Chern-Simons action $S_{\rm CS}$.


As we mentioned in the introduction, however,
it may be possible that the symmetry enhances with the decoupling
of $S_{\rm YM}$ and an appropriate choice of $k_I$.
Indeed, it is
shown in \cite{Hosomichi:2008jd} that
if the Chern-Simons coupling
is given by (\ref{ki}) with
\begin{equation}
s_I=(-1)^I,
\label{alt}
\end{equation}
the R-symmetry $SU(2)_D$ enhances to
$SU(2)\times SU(2)$.
We should note that this enhanced symmetry
acts on component fields in a different way from the original $SU(2)_L\times SU(2)_R$
symmetry.
We denote the new symmetry by $R_{\rm CS}=SU(2)_{+1}\times SU(2)_{-1}$.
In the model with (\ref{alt}),
the component fields in the hyper multiplets
belongs to the representation shown in Table \ref{hyprep.tbl}\cite{Hosomichi:2008jd}.
\begin{table}[htb]
\caption{
$R_{\rm CS}=SU(2)_{+1}\times SU(2)_{-1}$
representations of component fields
of hypermultiplets are shown.}
\label{hyprep.tbl}
\begin{center}
\begin{tabular}{c|cc}
\hline
\hline
                   & $q_I$ & $\psi_I$ \\
\hline
$s_I=1$  & $({\bf2},{\bf1})$ & $({\bf1},{\bf2})$ \\
$s_I=-1$ & $({\bf1},{\bf2})$ & $({\bf2},{\bf1})$ \\
\hline
\end{tabular}
\end{center}
\end{table}
A hypermultiplet $(q_I,\psi_I)$ with $s_I=1$ is
transformed in a different way from
a multiplet with $s_I=-1$.
These two types of hypermultiplets with different $s_I$ are
called hyper and twisted hyper multiplet in \cite{Hosomichi:2008jd}.
In the following
we prove $R_{\rm CS}$ invariance of our model based on the assumption
that
$(q_I,\psi_I)$ are transformed in the same way even when $s_I$ are not given by (\ref{alt}).

In order to show the enhancement of R-symmetry,
we integrate out $\lambda_I$ and $F_I$ in dynamical
vector multiplets.
The equation of motion of $F_I$ is
\begin{equation}
\frac{k_I}{2}\phi_I^A{}_B
=\mu_I^A{}_B-\wt\mu_{I-1}^A{}_B,
\label{feom}
\end{equation}
and we can eliminate the $\phi_I$ component of
the dynamical vector multiplet.
At the same time, $F_I$ itself disappears from the action.
The equation of motion of $\lambda_I$ is
\begin{equation}
k_I\lambda_I^{BA}
=
j_I^{AB}-\wt j_{I-1}^{AB}.
\label{lambdaeom}
\end{equation}
We eliminate $\lambda_I$
in the dynamical vector multiplet by this equation.

The resulting action includes
the following fields
\begin{equation}
\left\{\begin{array}{l}
\mbox{$(q_I,\psi_I)$ in hyper multiplets}\\
\mbox{$(v_{I\mu})$ in dynamical vector multiplets}\\
\mbox{$(v_{I\mu},\phi_I,\lambda_I,F_I)$ in auxiliary vector multiplets}
\end{array}\right.
\label{contents}
\end{equation}

\section{${\cal N}=4$ supersymmetry transformation}\label{susy.sec}
\subsection{Hyper multiplets}\label{hsusy.sec}
Now let us write down the ${\cal N}=4_{\rm CS}$ supersymmetry transformation.
This is achieved by rewriting
${\cal N}=3$ transformation in $R_{\rm CS}$ covariant form.

${\cal N}=3$ transformation is obtained from
that of ${\cal N}=4_{\rm YM}$ given in the previous section
by neglecting the distinction
between undotted and dotted indices,
and make the transformation parameter $\xi_{AB}$ symmetric
with respect to the exchange of two $SU(2)$ indices.

From this ${\cal N}=3$ transformation,
we can obtain ${\cal N}=4_{\rm CS}$ transformation
by carefully introducing distinction
between $SU(2)_{+1}$ and $SU(2)_{-1}$ indices
so that
$q_I$ and $\psi_I$ belongs to
the representations shown in
Table \ref{hyprep.tbl},
and
indices are contracted
among the same kind of indices.
We use overlined and underlined indices for
$SU(2)_{+1}$ and $SU(2)_{-1}$, respectively.
Two indices of the parameter $\xi$
are associated with different $SU(2)$ in $R_{\rm CS}$.
We assume that
the first and the second index
are acted by $SU(2)_{+1}$ and $SU(2)_{-1}$, respectively.

Let us rewrite the transformation of
$q_I$ in (\ref{deltaq}) in the $R_{\rm CS}$ covariant form.
The $R_{\rm CS}$ representations of $q_I$ and $\psi_I$
depend on $s_I$,
and the contraction of indices
in the supersymmetry transformation
also depends on $s_I$.
\begin{equation}
\delta q_I^{\overline A}=
\sqrt{2}i(\xi^{\overline A\underline B}\psi_{I\underline B})
\quad(s_I=+1),\quad
\delta q_I^{\underline A}=
\sqrt{2}i(\xi^{\overline B\underline A}\psi_{I\overline B})
\quad(s_I=-1).
\label{deltaqcov}
\end{equation}
In the left and right transformations
in (\ref{deltaqcov}),
$SU(2)$ index of $\psi$ is contracted with the second and the first
index of $\xi$, respectively.

In general, if we have supersymmetry transformation
laws for $s_I=+1$,
we can always rewrite them into transformation laws for $s_I=-1$
by replacing overlined and underlined indices by underlined
and overlined ones, respectively,
and exchanging two indices of the parameter $\xi$.
In the following we give only transformation laws
for $s_I=+1$.

Let us consider the transformation
law of $\psi_{I\ul A}$.
The transformation (\ref{deltapsi}) includes
$\phi_I$ and $\phi_{I+1}$,
and we treat these fields in different ways
depending on $k_I$ and $k_{I+1}$.
If $k_I=0$ ($k_{I+1}$=0) we eliminate $\phi_I$ ($\phi_{I+1}$) by using
(\ref{feom}) while we leave it in the action if $k_I\neq0$ ($k_{I+1}\neq0$).
For example,
if $k_I=0$ and $k_{I+1}\neq0$
we leave $\phi_I$ in the action and
eliminate $\phi_{I+1}$ by (\ref{feom}).
From (\ref{deltapsi}) we obtain ${\cal N}=3$
transformation as
\begin{equation}
\delta\psi_{I\ul A}
=
\sqrt2\xi_{CB}\phi_I^B{}_Aq_I^C
+\frac{2s_I}{k}\sqrt2\xi_{CB}q_I^C\wt\mu_I^B{}_A
-\frac{2s_I}{k}\sqrt2\xi_{\ol C\ul B}q_I^{\ol C}\mu_{I+1}^{\ul B}{}_{\ul A}
+\sqrt2\gamma^\mu\xi_{\ol B\ul A} D_\mu q_I^{\ol B}.
\label{deltapsix}
\end{equation}
We put overlines and underlines to the indices in
the third and fourth terms.
However, it is impossible to do it consistently
in the second term.

In order to resolve this problem
we introduce the following shifted field.
\begin{equation}
\varphi_I^A{}_B
=\phi_I^A{}_B-\frac{s_I}{k}(\mu_I^A{}_B+\wt\mu_{I-1}^A{}_B).
\label{phishift}
\end{equation}
By this field redefinition
we rewrite the transformation (\ref{deltapsix}) for general $k_I$ and $k_{I+1}$
as
\begin{eqnarray}
\delta\psi_{I\ul A}
&=&
\sqrt2\gamma^\mu\xi_{\ol B\ul A}D_\mu q_I^{\ol B}
-\frac{\sqrt2s_I}{k}\xi_{\ol C\ul A}(\nu_I^{\ol D}{}_{\ol D}q_I^{\ol C}-q_I^{\ol C}\wt\nu_I^{\ol D}{}_{\ol D})
\nonumber\\&&
+\left(\sqrt2\xi_{\ol C\ul B}\varphi_I^{\ul B}{}_{\ul A}q_I^{\ol C}\right)_{k_I=0}
-\left(\frac{2\sqrt2s_I}{k}\xi_{\ol C\ul B}\wt\mu_{I-1}^{\ul B}{}_{\ul A}q_I^{\ol C}\right)_{k_I\neq0}
\nonumber\\&&
-\left(\sqrt2\xi_{\ol C\ul B}q_I^{\ol C}\varphi_{I+1}^{\ul B}{}_{\ul A}\right)_{k_{I+1}=0}
+\left(\frac{2\sqrt2s_I}{k}\xi_{\ol C\ul B}q_I^{\ol C}\mu_{I+1}^{\ul B}{}_{\ul A}\right)_{k_{I+1}\neq0}
\nonumber\\&&
+\delta'\psi_{I\ol A},
\label{deltapsicov}
\end{eqnarray}
where $(\cdots)_{\rm condition}$ means that it is included only when the condition
is satisfied.
This transformation still includes non-covariant terms
and we collected them into the last term,
$\delta'\psi_{I\ol A}$, which is given by
\begin{equation}
\delta'\psi_{I\ol A}
=-\left(\frac{\sqrt2s_I}{k}\xi_{CB}(\mu_I-\wt\mu_{I-1})^B{}_Aq_I^C\right)_{k_I=0}
-\left(\frac{\sqrt2s_I}{k}\xi_{CB}q_I^C(\mu_{I+1}-\wt\mu_I)^B{}_A\right)_{k_{I+1}=0}.
\end{equation}
We will comment on this non-covariant part at the end of the next subsection.
It will there be turn out that we can easily remove this unwanted part
from the transformation law.

\subsection{Vector multiplets}\label{vsusy.sec}
Let us write down the ${\cal N}=4_{\rm CS}$ transformation law
for vector multiplets.
If a vector multiplet is dynamical,
it has only one component $v_{I\mu}$ as shown in (\ref{contents}),
and by using (\ref{lambdaeom}) the transformation law
(\ref{deltav0}) is rewritten as
\begin{equation}
\delta v_{I\mu}
=-\frac{s_I}{k}
\xi_{\overline A\underline B}\gamma_\mu
(
j_I^{\overline A\underline B}
-\wt j_{I-1}^{\underline B\overline A}).
\label{deltadvcov}
\end{equation}
This is $R_{\rm CS}$ invariant.

In an auxiliary vector multiplet,
we have four component fields.
In order to write manifestly $R_{\rm CS}$ covariant
${\cal N}=4_{\rm CS}$ transformation laws,
we need to shift the fields $\lambda_I$ and $F_I$
as well as $\phi$ in the following way.
\begin{eqnarray}
\lambda_I'^{AB}
&=&\lambda_I^{AB}-\frac{s_I}{2k}(j_I^{BA}+\wt j_{I-1}^{BA})
\label{lambdashift},\\
F_I'^A{}_B
&=&F_I^A{}_B+\frac{s_I}{k}(K_I^A{}_B+\wt K_{I-1}^A{}_B)
\nonumber\\&&
+\frac{s_I}{2k}[(\mu_I+\wt\mu_{I-1})^A{}_C,\varphi_I^C{}_B]
-\frac{s_I}{2k}[(\mu_I+\wt\mu_{I-1})^C{}_B,\varphi_I^A{}_C],
\label{Fshift}
\end{eqnarray}
where $K_I$ and $\wt K_I$ in (\ref{Fshift})
and $J_I^\mu$ and $\wt J_I^\mu$ appearing in (\ref{deltava}) below
are components of current multiplets
defined in Appendix \ref{current.sec}.
The transformation laws of $v_{I\mu}$, $\varphi_I$,
and $\lambda_I'$ are manifestly covariant.
\begin{eqnarray}
\delta v_{I\mu}
&=&
-\xi_{\ol A\ul B}\gamma_\mu\left(\lambda_I'^{\ol A\ul B}+\frac{s_I}{2k}(j_I^{\ol A\ul B}+\wt j_{I-1}^{\ol A\ul B})\right),
\label{deltava}\\
\delta\varphi_I^{\ul A}{}_{\ul B}
&=&2i\xi_{\ol C\ul B}\lambda_I'^{\ol C\ul A}
-i\delta^{\ul A}{}_{\ul B}\xi_{\ol C\ul D}\lambda_I'^{\ol C\ul D},
\label{lambdap}\\
\delta\lambda_I'^{\ol A\ul B}
&=&\frac{i}{2}\gamma^{\mu\nu}\xi^{\ol A\ul B} F_{I\mu\nu}
+\frac{is_I}{2k}\gamma_\mu\xi^{\ol A\ul B}(J_I^\mu+\wt J_{I-1}^\mu)
+\gamma^\mu\xi^{\ol A\ul C} D_\mu\varphi_I^{\ul B}{}_{\ul C}
+\xi^{\ol C\ul B}F_I'^{\ol A}{}_{\ol C}
\nonumber\\
&&
+\frac{1}{2}[\varphi_I^{\ul B}{}_{\ul C},\varphi_I^{\ul C}{}_{\ul D}]\xi^{\ol A\ul D}
+\frac{1}{2k^2}[(\mu_I+\wt\mu_{I-1})^{\ol A}{}_{\ol C},(\mu_I+\wt\mu_{I-1})^{\ol C}{}_{\ol D}]
\xi^{\ol D\ul B}.\label{deltalambdacov}
\end{eqnarray}
The transformation of $F_I'^A{}_B$ includes non-covariant terms.
\begin{eqnarray}
\delta F_I'^{\ol A}{}_{\ol B}
&=&
2i\xi_{\ol B\ul C}\gamma^\mu D_\mu\lambda_I'^{\ol A\ul C}
+2i\xi_{\ol B\ul C}[\lambda_I'^{\ol A\ul D},\varphi_I^{\ul C}{}_{\ul D}]
\nonumber\\&&
+\frac{is_I}{k}[\xi_{\ol B\ul C}(j_I+\wt j_{I-1})^{\ol A\ul D},\varphi_I^{\ul C}{}_{\ul D}]
\nonumber\\&&
-\frac{2is_I}{k}[\xi_{\ol B\ul D}\lambda_I'^{\ol C\ul D}-\tr,(\mu_I+\wt\mu_{I-1})^{\ol A}{}_{\ol C}]
\nonumber\\&&
+\frac{i}{k^2}[\xi_{\ol B\ul D}(j_I+\wt j_{I-1})^{\ol C\ul D}-\tr,(\mu_I+\wt\mu_{I-1})_I^{\ol A}{}_{\ol C}]
\nonumber\\&&
+\delta'F_I'^{\ol A}{}_{\ol B}.\label{deltafcov}
\end{eqnarray}
We collected non-covariant terms
into $\delta' F'_I$.
It is given by
\begin{equation}
\delta'F_I'^{\ol A}{}_{\ol B}
=\frac{\sqrt2is_I}{k}\xi_{CB}(q_I^C\ol\Psi_I^A+\ol\Psi_{I-1}^Aq_{I-1}^C)
+\frac{\sqrt2is_I}{k}\xi^{CA}(\Psi_{IB}\ol q_{IC}+\ol q_{I-1C}\Psi_{I-1B}),
\end{equation}
where $\Psi_{IA}$ is the left hand side of
the equation of motion $\Psi_{IA}=0$ of the fermion $\psi_{IA}$.
\begin{equation}
\Psi_{IA}
=\gamma^\mu D_\mu\psi_{IA}
-\phi_I^B{}_A\psi_{IB}
+\psi_{IB}\phi_{I+1}^B{}_A
+\sqrt2\lambda_{IBA}q_I^B
-\sqrt2q_I^B\lambda_{I+1BA}.
\label{psieom}
\end{equation}

Among the supersymmetry transformation laws
written down in the previous and this subsections,
$\delta\psi_I$ and $\delta F_I'$ include non-covariant parts
$\delta'\psi_I$ and $\delta' F'_I$.
These non-covariant terms may be simply removed
from the transformation
because, as is easily checked, the action $S_{\rm CS}+S_{\rm hyper}$ is in fact invariant under the non-covariant transformation $\delta'$.
Removing these terms,
we obtain completely $R_{\rm CS}$ covariant ${\cal N}=4_{\rm CS}$
supersymmetry transformation laws.

\section{$SU(2)\times SU(2)$ invariance of the action}\label{action.sec}
In this section, we prove the $R_{\rm CS}$ invariance
of the action $S_{\rm CS}+S_{\rm hyper}$.
Here
we use ${\cal N}=3$ notation to simplify equations.
Namely, we use plain indices without dots or lines for any $SU(2)$.
It is easy to check if each term is
$R_{\rm CS}$ invariant or not.

We first rearrange
the action into the following three parts.
The first part, $\wh S_{\rm kin}$, includes the kinetic terms.
\begin{eqnarray}
\wh S_{\rm kin}&=&
\sum_{I=1}^n
\int d^3x\tr\Big[
k_I
\epsilon^{\mu\nu\rho}
\left(\frac{1}{2}v_{I\mu}\partial_\nu v_{I\rho}-\frac{i}{3}v_{I\mu} v_{I\nu} v_{I\rho}\right)
\nonumber\\&&
-D_\mu\ol q_{IA} D^\mu q_I^A
-i\ol\psi_I^A\gamma^\mu D_\mu\psi_{IA}
\Big].
\label{Skin}
\end{eqnarray}
This part is manifestly $R_{\rm CS}$ invariant.
We use hats for manifestly $R_{\rm CS}$ invariant terms.

The second part, $S_{\rm pot}$, includes potential terms
\begin{eqnarray}
S_{\rm pot}&=&
\sum_I\int d^3x\tr\Big[
\frac{k_I}{2}\phi_I^A{}_BF_I^B{}_A
-F_I^A{}_B(\mu_I^B{}_A-\wt\mu_{I-1}^B{}_A)
\nonumber\\&&
-\frac{1}{2}\nu_I^A{}_A\phi_I^B{}_C\phi_I^C{}_B
-\frac{1}{2}\wt\nu_I^A{}_A\phi_{I+1}^B{}_C\phi_{I+1}^C{}_B
+\frac{k_I}{6}\phi_I^A{}_B\phi_I^B{}_C\phi_I^C{}_A
\nonumber\\&&
+\ol q_{IA}\phi_I^B{}_Cq_I^A\phi_{I+1}^C{}_B
\Big].
\label{Spot}
\end{eqnarray}
This part is analyzed in \S\ref{pot.sec}.

The rest of the action is the following part including Yukawa terms.
\begin{eqnarray}
S_{\rm Yukawa}
&=&
\sum_I
\int d^3x
\tr\Big[
\frac{ik_I}{2}\lambda_I^{AB}\lambda_{IBA}
-i\lambda_{IAB}(j_I^{AB}-\wt j_{I-1}^{AB})
\nonumber\\&&
+i\psi_{IB}\ol\psi_I^A\phi_I^B{}_A
-i\ol\psi_{I-1}^A\psi_{I-1B}\phi_I^B{}_A
\Big].
\label{Syukawa}
\end{eqnarray}
This part is analyzed in \S\ref{yukawa.sec}.

\subsection{Potential terms}\label{pot.sec}
We decompose the potential term by
\begin{equation}
S_{\rm pot}=\sum_{I=1}^n(S_{\rm pot1}^{I(k_I)}+S_{\rm pot2}^{I(k_I,k_{I+1})}),
\end{equation}
where $S_{\rm pot1}^{I(k_I)}$ and $S_{\rm pot 2}^{I(k_I,k_{I+1})}$ are defined by
\begin{eqnarray}
S_{\rm pot1}^{I(k_I)}&=&
\int d^3x\tr\Big[
\frac{k_I}{2}\phi_I^A{}_BF_I^B{}_A
-F_I^A{}_B(\mu_I^B{}_A-\wt\mu_{I-1}^B{}_A)
\nonumber\\&&
-\frac{1}{2}\nu_I^A{}_A\phi_I^B{}_C\phi_I^C{}_B
-\frac{1}{2}\wt\nu_{I-1}^A{}_A\phi_I^B{}_C\phi_I^C{}_B
+\frac{k_I}{6}\phi_I^A{}_B\phi_I^B{}_C\phi_I^C{}_A
\Big],\\
S_{\rm pot2}^{I(k_I,k_{I+1})}&=&
\int d^3x\tr(\ol q_{IA}\phi_I^B{}_Cq_I^A\phi_{I+1}^C{}_B).
\end{eqnarray}
$S_{\rm pot1}^{I(k_I)}$ includes only one $\phi_I$
while $S_{\rm pot2}^{I(k_I,k_{I+1})}$ includes $\phi_I$ and $\phi_{I+1}$.

We first consider $S_{\rm pot1}^{I(k_I)}$.
When $k_I\neq0$, we eliminate $\phi_I$ by using (\ref{feom}).
Then $S_{\rm pot1}^{I(k_I)}$ includes
only scalar fields $q_I$, $q_{I-1}$, and their Hermitian conjugates.
\begin{eqnarray}
S_{\rm pot1}^{I(k_I\neq0)}
&=&
\int d^3x\tr\left[
\frac{4}{k^2}q_I^A\wt\mu_I^B{}_C\ol q_{IA}\wt\mu_{I-1}^C{}_B
+\frac{4}{k^2}\ol q_{I-1B}\mu_{I-1}^A{}_Cq_{I-1}^B\mu_I^C{}_A
\right]
\nonumber\\&&
+\wh S_{\rm pot1}^{I(k_I\neq0)},
\label{Spot11}\\
\wh S_{\rm pot1}^{I(k_I\neq0)}
&=&
\frac{2}{k^2}\int d^3x\tr\Big[
-\mu_I^A{}_B\mu_I^B{}_A\wt\nu_{I-1}^C{}_C
-\wt\mu_{I-1}^A{}_B\wt\mu_{I-1}^B{}_A\nu_I^C{}_C
\nonumber\\&&\hspace{3em}
-\nu_I^A{}_A\mu_I^B{}_C\mu_I^C{}_B
-\wt\nu_{I-1}^A{}_A\wt\mu_{I-1}^B{}_C\wt\mu_{I-1}^C{}_B
\nonumber\\&&\hspace{3em}
+\frac{2}{3}\mu_I^A{}_B\mu_I^B{}_C\mu_I^C{}_A
-\frac{2}{3}\wt\mu_{I-1}^A{}_B\wt\mu_{I-1}^B{}_C\wt\mu_{I-1}^C{}_A
\Big].
\end{eqnarray}
Because we now assume $k_I\neq0$,
$q_I$ and $q_{I-1}$ are transformed by
different $SU(2)$ factors in $R_{\rm CS}$.
Thus, if $SU(2)$ indices of $q_I$ and those of $q_{I-1}$ are contracted,
the term breaks the $R_{\rm CS}$ symmetry.
To prove the $R_{\rm CS}$ invariance of the action,
we need to show that such terms cancel among them
when we sum up all terms in the action.
By this reason, we separate manifestly $R_{\rm CS}$ invariant
terms and denote them by $\wh S_{\rm pot1}^{I(k_I)}$.
In each term in $\wh S_{\rm pot1}^{I(k_I)}$
indices of $q_I$ and those of $q_{I-1}$ are separately contracted.
Contrary, in the first line of (\ref{Spot11})
some indices of $q_I$ are contracted with $q_{I-1}$, and
breaks the $R_{\rm CS}$ symmetry.

When $k_I=0$,
we rewrite the field $\phi_I$ and $F_I$
by the $R_{\rm CS}$ covariant field $\varphi_I$ and $F'_I$
defined in \S\ref{susy.sec}.
We obtain
\begin{equation}
S_{\rm pot 1}^{I(k_I=0)}=
\int d^3x\tr\left[
-\frac{2s_I}{k}\wt\nu_{I-1}^A{}_B\wt\nu_{I-1}^B{}_C\varphi_I^C{}_A
-\frac{2s_I}{k}\nu_I^A{}_C\nu_I^B{}_A\varphi_I^C{}_B
\right]
+\wh S_{\rm pot 1}^{I(k_I=0)}
+C^I,
\label{Spot12}
\end{equation}
where we collected $R_{\rm CS}$ invariant terms into $\wh S_{\rm pot1}^{I(k_I=0)}$
\begin{eqnarray}
\wh S_{\rm pot 1}^{I(k_I=0)}
&=&
\int d^3x\tr\Big[
-\frac{1}{2}\nu_I^A{}_A\varphi_I^B{}_C\varphi_I^C{}_B
-\frac{2}{k^2}\nu_I^A{}_A\mu_I^B{}_C\mu_I^C{}_B
\nonumber\\&&
-\frac{1}{2}\wt\nu_{I-1}^A{}_A\varphi_I^B{}_C\varphi_I^C{}_B
-\frac{2}{k^2}\wt\nu_{I-1}^A{}_A\wt\mu_{I-1}^B{}_C\wt\mu_{I-1}^C{}_B
\nonumber\\&&
-F_I'^A{}_B(\mu_I-\wt\mu_{I-1})^B{}_A
\nonumber\\&&
+\frac{1}{2k^2}(\nu_I+\wt\nu_{I-1})^A{}_A
(\mu_I-\wt\mu_{I-1})^B{}_C(\mu_I-\wt\mu_{I-1})^C{}_B
\Big],
\end{eqnarray}
and $C^I$ is defined by
\begin{equation}
C^I=\frac{is}{k}(\psi_{IA}\ol\psi_I^B+\ol\psi_{I-1}^B\psi_{I-1A})(\mu_I^A{}_B-\wt\mu_{I-1}^A{}_B).
\label{cidef}
\end{equation}

It is convenient to write (\ref{Spot11}) and (\ref{Spot12}) in the unified form
\begin{equation}
S_{\rm pot1}^{I(k_I)}
=B^{I(k_I)}+A^{I(k_I)}+\wh S_{\rm pot1}^{I(k_I)}+(C^I)_{k_I=0},
\label{Spot1x}
\end{equation}
where
$A^{I(k_I)}$ and $B^{I(k_I)}$ are defined by
\begin{eqnarray}
A^{I(k_I\neq0)}
&=&\frac{4}{k^2}\int d^3x\tr
(q_I^A\wt\mu_I^B{}_C\ol q_{IA}\wt\mu_{I-1}^C{}_B),\\
A^{I(k_I=0)}
&=&\frac{s_I}{k}\int d^3x
\tr(-2\nu_I^A{}_C\nu_I^B{}_A\varphi_I^C{}_B
+q_I^C\phi_{I+1}^A{}_B\ol q_{IC}(\mu_I-\wt\mu_{I-1})^B{}_A)
,\\
B^{I(k_I\neq0)}
&=&\frac{4}{k^2}\int d^3x
\tr(\ol q_{I-1B}\mu_{I-1}^A{}_Cq_{I-1}^B\mu_I^C{}_A),\\
B^{I(k_I=0)}
&=&\frac{s_I}{k}\int d^3x
\tr(-2\wt\nu_{I-1}^A{}_B\wt\nu_{I-1}^B{}_C\varphi_I^C{}_A
-\ol q_{I-1C}\phi_{I-1}^A{}_Bq_{I-1}^C(\mu_I-\wt\mu_{I-1})^B{}_A).
\end{eqnarray}

Next, let us consider $S_{\rm pot2}^{I(k_I,k_{I+1})}$.
This term contains $\phi_I$ and $\phi_{I+1}$,
and we need to consider four cases separately according to whether $k_I$ and $k_{I+1}$ are zero or not.
When $k_I\neq 0$, we use (\ref{feom}) to eliminate $\phi_I$,
and when $k_I=0$ we rewrite the field $\phi_I$ according to (\ref{phishift}).
We treat $\phi_{I+1}$ in the same way, too.
The result is
\begin{equation}
S_{\rm pot2}^{I(k_I,k_{I+1})}
=-A^{I(k_I)}-B^{I+1(k_{I+1})}+\wh S_{\rm pot2}^{I(k_I,k_{I+1})}.
\label{Spot2x}
\end{equation}
We collected manifestly $R_{\rm CS}$ invariant terms
into $\wh S_{\rm pot2}^{I(k_I,k_{I+1})}$.
It is given by
\begin{eqnarray}
\wh S_{\rm pot2}^{I(k_I\neq0,k_{I+1}\neq0)}
&=&
\int d^3x\tr\left[
\frac{4}{k^2}\ol q_{IA}\mu_I^B{}_Cq_I^A\wt\mu_I^C{}_B
+\frac{4}{k^2}\ol q_{IA}\wt\mu_{I-1}^B{}_Cq_I^A\mu_{I+1}^C{}_B
\right],\\
\wh S_{\rm pot2}^{I(k_I\neq0,k_{I+1}=0)}
&=&
\int d^3x\tr\left[
-\frac{4}{k^2}\ol q_{IA}\mu_I^B{}_Cq_I^A\wt\mu_I^C{}_B
-\frac{2s_I}{k}\ol q_{IA}\wt\mu_{I-1}^B{}_Cq_I^A\varphi_{I+1}^C{}_B
\right],\\
\wh S_{\rm pot2}^{I(k_I=0,k_{I+1}\neq0)}
&=&
\int d^3x\tr\left[
-\frac{2s_I}{k}\ol q_{IA}\varphi_I^B{}_Cq_I^A\mu_{I+1}^C{}_B
 +\frac{4}{k^2}\ol q_{IA}\mu_I^B{}_Cq_I^A\wt\mu_I^C{}_B
\right],\\
\wh S_{\rm pot2}^{I(k_I=0,k_{I+1}=0)}
&=&
\int d^3x\tr\Big[
\ol q_{IA}\varphi_I^B{}_Cq_I^A\varphi_{I+1}^C{}_B
-\frac{4}{k^2}\ol q_{IA}\mu_I^B{}_Cq_I^A\wt\mu_I^C{}_B
\nonumber\\&&\hspace{3em}
+\frac{1}{k^2}\ol q_{IC}(\mu_I-\wt\mu_{I-1})^A{}_Bq_I^C(\mu_{I+1}-\wt\mu)^B{}_A
\Big].
\end{eqnarray}

If we sum up (\ref{Spot1x}) and (\ref{Spot2x}) over all $I$,
all $A^{I(k_I)}$ and $B^{I(k_I)}$ cancel
and we obtain
\begin{equation}
S_{\rm pot}=\sum_{I=1}^n(\wh S_{\rm pot1}^{I(k_I)}+\wh S_{\rm pot2}^{I(k_I,k_{I+1})})
+\sum_{k_I=0}C^I.
\label{Spotfin}
\end{equation}

\subsection{Yukawa terms}\label{yukawa.sec}
Let us consider $S_{\rm Yukawa}$ in (\ref{Syukawa}).
We decompose it as
\begin{equation}
S_{\rm Yukawa}=\sum_{I=1}^n S_{\rm Yukawa}^{I(k_I)},
\end{equation}
where 
\begin{eqnarray}
S_{\rm Yukawa}^{I(k_I)}
&=&
\int d^3x\tr\Big[
\frac{ik_I}{2}\lambda_I^{AB}\lambda_{IBA}
-i\lambda_{IAB}(j_I^{AB}-\wt j_{I-1}^{AB})
\nonumber\\&&
+i\psi_{IB}\ol\psi_I^A\phi_I^B{}_A
-i\ol\psi_{I-1}^A\psi_{I-1B}\phi_I^B{}_A
\Big].
\end{eqnarray}
Again we should discuss two cases with $k_I\neq0$ and $k_I=0$ separately.

If $k_I\neq0$,
eliminating $\lambda_I$ by using the equation of motion
(\ref{lambdaeom}),
and rewriting $\phi_I$ by (\ref{feom}),
we obtain
\begin{equation}
S_{\rm Yukawa}^{I(k_I\neq0)}
=\frac{i}{k_I}(Y_{I-1}+X_I)+\wh S_{\rm Yukawa}^{I(k_I\neq0)},
\label{syn0}
\end{equation}
where we defined
\begin{eqnarray}
X_I&=&
\int d^3x\tr\left[
-\frac{1}{2}j_I^{AB}j_{IBA}
+2\psi_{IB}\ol\psi_I^A\mu_I^B{}_A\right],\\
Y_I&=&
\int d^3x\tr\left[
-\frac{1}{2}\wt j_I^{AB}\wt j_{IBA}
+2\ol\psi_I^A\psi_{IB}\wt\mu_I^B{}_A\right],
\end{eqnarray}
and
\begin{equation}
\wh S_{\rm Yukawa}^{I(k_I\neq0)}
=
\frac{i}{k_I}\int d^3x\tr\Big[
\wt j_{I-1BA}j_I^{AB}
-2\psi_{IB}\ol\psi_I^A\wt\mu_{I-1}^B{}_A
-2\ol\psi_{I-1}^A\psi_{I-1B}\mu_I^B{}_A
\Big].
\end{equation}
When $k_I\neq 0$,
$q_I$ and $\psi_I$ are rotated by
the same $SU(2)$ as $\psi_{I-1}$ and $q_{I-1}$,
respectively,
and we see that terms in
$\wh S_{\rm Yukawa}^{I(k_I\neq0)}$ are
manifestly $R_{\rm CS}$ invariant while
$X$ and $Y$ are not.
We define
\begin{eqnarray}
\wh Z_I
&=&\int d^3x\tr[
\epsilon_{AB}\epsilon_{CD}q_I^A\ol\psi_I^Cq_I^B\ol\psi_I^D
-\epsilon^{AB}\epsilon^{CD}\ol q_{IA}\psi_{IC}\ol q_{IB}\psi_{ID}
\nonumber\\&&\hspace{10em}
+\psi_{IA}\ol\psi_I^Aq_I^B\ol q_{IB}
-\ol\psi_I^A\psi_{IA}\ol q_{IB}q_I^B
].
\end{eqnarray}
This is manifestly $R_{\rm SC}$ invariant, and
the following identity holds.
\begin{equation}
Y_I-X_I=\wh Z_I.
\label{keyeq}
\end{equation}
By using this identity, we can rewrite the action (\ref{syn0}) as
\begin{equation}
S_{\rm Yukawa}^{I(k_I\neq0)}
=
\frac{i}{k}
\Big[
-s_{I-1}X_{I-1}
+s_IX_I
\Big]
-\frac{is_{I-1}}{k}Z_{I-1}
+\wh S_{\rm Yukawa}^{I(k_I\neq0)},
\label{Sy1}
\end{equation}
where we used the relation $s_I=-s_{I-1}$, which holds when $k_I\neq0$.

Next, let us consider $k_I=0$ case.
Rewriting $\phi_I$ and $\lambda_I$ in the action according to
(\ref{phishift}) and (\ref{lambdashift})
we obtain
\begin{eqnarray}
S_{\rm Yukawa}^{I(k_I=0)}
&=&
\frac{is_I}{k}(-Y_{I-1}+X_I)
+\wh S_{\rm Yukawa}^{I(k_I=0)}
\nonumber\\
&=&
\frac{i}{k}(-s_{I-1}X_{I-1}+s_IX_I)
-\frac{is_{I-1}}{k}\wh Z_{I-1}
+\wh S_{\rm Yukawa}^{I(k_I=0)}
-C^I,
\label{Sy0}
\end{eqnarray}
where $C^I$ is defined in (\ref{cidef}),
and $\wh S_{\rm Yukawa}^{I(k_I=0)}$ includes $R_{\rm CS}$ invariant terms.
\begin{equation}
\wh S_{\rm Yukawa}^{I(k_I=0)}
=
\int d^3x\tr\Big[
-i\ol\psi_{I-1}^A\psi_{I-1B}\varphi_I^B{}_A
+i\psi_{IB}\ol\psi_I^A\varphi_I^B{}_A
-i\lambda'_{IAB}(j_I^{AB}-\wt j_{I-1}^{AB})
\Big].
\end{equation}

Summing up $S_{\rm Yukawa}^{I(k_I)}$
in (\ref{Sy1}) and (\ref{Sy0})
over all $I$, terms with $X_I$ and $Y_I$ cancel,
and we obtain
\begin{equation}
S_{\rm Yukawa}
=\sum_{I=1}^n\left(
-\frac{is_I}k\wh Z_I+\wh S_{\rm Yukawa}^{I(k_I)}
\right)
-\sum_{k_I=0}C^I.
\label{Syukfin}
\end{equation}

Adding (\ref{Spotfin}) and (\ref{Syukfin}),
we obtain the manifestly $R_{\rm CS}$ invariant action
\begin{equation}
S_{\rm CS}+S_{\rm hyper}=\wh S_{\rm kin}
+\sum_{I=1}^n\left(
\wh S_{\rm pot1}^{I(k_I)}
+\wh S_{\rm pot2}^{I(k_I,k_{I+1})}
-\frac{is_I}k\wh Z_I+\wh S_{\rm Yukawa}^{I(k_I)}
\right),
\label{shat}
\end{equation}
and the proof is completed.

\section{Conclusions}\label{conc.sec}
In this paper we investigated the $Spin(4)$ R-symmetry
and ${\cal N}=4$ supersymmetry of the three-dimensional
Chern-Simons-matter system described by
the action $S_{\rm CS}+S_{\rm hyper}$,
where $S_{\rm CS}$ and $S_{\rm hyper}$
are given in (\ref{ScsSF}) and (\ref{ShyperSF}), respectively.
This model consists of dynamical and auxiliary vector multiplets
and bi-fundamental hypermultiplets.
The dynamical vector multiplets have Chern-Simons couplings $\pm k$ while
the auxiliary vector multiplets do not have Chern-Simons terms.
(Although we call vector multiplets with non-vanishing Chern-Simons couplings
``dynamical'' for distinction, they do not have propagating degrees of freedom.)
After integrating out auxiliary fields in the hyper and dynamical vector multiplets,
our model includes
$(q_I,\psi_I)$ in the hypermultiplets,
$(v_{I\mu})$ in the dynamical vector multiplets,
and $(v_{I\mu},\varphi_I,\lambda_I',F_I')$ in the auxiliary vector multiplets.
We wrote down the ${\cal N}=4$ supersymmetry transformation
in terms of these component fields
in manifestly $Spin(4)$ covariant form in
Eqs. (\ref{deltaqcov}), (\ref{deltapsicov}), and (\ref{deltava}-\ref{deltafcov}).
We also proved the ${\cal N}=4$ invariance of the action in \S\ref{action.sec}
by rewriting it in the manifestly $Spin(4)$ invariant form (\ref{shat}).

\section*{Acknowledgements}
Y.~I. is partially supported by
Grant-in-Aid for Young Scientists (B) (\#19740122) from the Japan
Ministry of Education, Culture, Sports,
Science and Technology.

\appendix
\section{${\cal N}=4$ multiplets and ${\cal N}=2$ superfields}\label{sf.sec}
In this appendix we summarize our conventions
for spinors and superfields.
Because all we need in this paper
are actions and transformation laws in terms of component fields,
which are given in the main text,
we here do not present detail of the superfield formalism.
The purpose of this appendix is to show
rough relation between components
and superfields.

We use $(-++)$ signature for the metric,
and $\gamma^\mu$ are real $2\times 2$ matrices satisfying
\begin{equation}
\eta^{\mu\nu}=\frac{1}{2}\tr(\gamma^\mu\gamma^\nu),\quad
\epsilon^{\mu\nu\rho}
=\frac{1}{2}\tr(\gamma^\mu\gamma^\nu\gamma^\rho).
\end{equation}
To make fermion bi-linears, we use the antisymmetric tensor $\epsilon_{\alpha\beta}$
defined by
\begin{equation}
\epsilon_{12}=-\epsilon_{21}=1.
\end{equation}
For example,
\begin{equation}
(\eta\chi)=\eta^\alpha\epsilon_{\alpha\beta}\chi^\beta,\quad
(\eta\gamma^\mu\chi)=\eta^\alpha\epsilon_{\alpha\beta}(\gamma^\mu)^\beta{}_\gamma
\chi^\gamma.
\end{equation}

Let $(x^\mu,\theta^\alpha,\ol\theta^\alpha)$ be the ${\cal N}=2$ superspace.
$\ol\theta^\alpha$ is the complex conjugate of the complex spinor $\theta^\alpha$.
The complex conjugate of the product of two Grassmann variables $\alpha$ and $\beta$
is defined by
$(\alpha\beta)^*=\beta^*\alpha^*$.

A vector superfield in the Wess-Zumino gauge is expanded as
\begin{equation}
V(v_\mu,\sigma,\lambda,D)
=(\theta\gamma^\mu\ol\theta)v_\mu
 +i(\theta\ol\theta)\sigma
 +\theta^2(\ol\theta\ol\lambda)
 +\ol\theta^2(\theta\lambda)
 +\frac{1}{2}\theta^2\ol\theta^2D.
\end{equation}
The transformation laws of component fields are
\begin{eqnarray}
\delta\sigma&=&i(\xi\ol\lambda)+i(\ol\xi\lambda),\\
\delta v_\mu&=&(\xi\gamma_\mu\ol\lambda)-(\ol\xi\gamma_\mu\lambda),\\
\delta D
&=&
i(\xi\gamma^\mu D_\mu\ol\lambda)
+i(\ol\xi\gamma^\mu D_\mu\lambda)
+i(\xi[\sigma,\ol\lambda])
+i(\ol\xi[\sigma,\lambda]),\\
\delta\lambda&=&\frac{i}{2}\gamma^{\mu\nu}\xi F_{\mu\nu}
+\gamma^\mu\xi D_\mu\sigma
+D\xi.
\end{eqnarray}

We expand a chiral superfield as
\begin{equation}
\Phi(\phi,\psi,F)
=\phi
+\sqrt2i\theta\psi
+i\theta^2F
+\mbox{$\ol\theta$ dependent terms}.
\end{equation}
The supersymmetry transformation in the Wess-Zumino gauge is
\begin{eqnarray}
\delta\phi&=&\sqrt2i(\xi\psi),\\
\delta\psi&=&
\sqrt2\xi F
+\sqrt2\ol\xi \sigma\phi
+\sqrt2\gamma^\mu\ol\xi D_\mu\phi,\\
\delta F
&=&
\sqrt2i(\ol\xi\gamma^\mu D_\mu\psi)
-\sqrt2i(\ol\xi\sigma\psi)
-2i(\ol\xi\ol\lambda)\phi.
\end{eqnarray}

An ${\cal N}=4$ vector multiplet is made of
an ${\cal N}=2$ vector multiplet $V$ with components $(v_\mu,\sigma,\lambda,D)$ and
an adjoint chiral multiplet $\Phi$ with components $(\phi,\chi,F_\phi)$.
In order to make the $R_{\rm YM}=Spin(4)$ symmetry manifest
we form the following $R_{\rm YM}$ multiplets.
\begin{equation}
\lambda^{A\dot B}
=
\left(\begin{array}{cc}
\lambda & \ol\chi \\
\chi & -\ol\lambda
\end{array}\right),\quad
\phi^{\dot A}{}_{\dot B}
=\left(\begin{array}{cc}
\sigma & \sqrt2\phi \\
\sqrt2\ol\phi & -\sigma
\end{array}\right),\quad
F^A{}_B
=\left(\begin{array}{cc}
D' &  \sqrt2\ol F_\phi \\
\sqrt2 F_\phi & -D'
\end{array}\right),
\label{fabdef}
\end{equation}
where $D'$ is the shifted auxiliary field
\begin{equation}
D'=D-[\phi,\ol\phi].
\end{equation}

A hypermultiplet is made of
two chiral multiplets
$Q(q,\psi,F)$
and
$\wt Q(\wt q,\wt\psi,\wt F)$.
These two chiral multiplets must belong to conjugate representations
of gauge group to each other.
We define the following $R_{\rm YM}$ doublets.
\begin{equation}
q^A=(q^1,q^2)=(q,\ol{\wt q}),\quad
\psi_{\dot A}=(\psi_{\dot1},\psi_{\dot2})=(\psi,\ol{\wt\psi}).
\end{equation}

\section{Current multiplets}\label{current.sec}
The components of current multiplets are defined by
the differentiation of the action $S_{\rm hyper}$ given in (\ref{Shyper})
with respect to the components of vector multiplets.
\begin{eqnarray}
\delta S_{\rm hyper}^I
&=&
-\delta F_I^A{}_B\mu_I^B{}_A
-i\delta\lambda_{IA\dot B}j_I^{A\dot B}
+\delta v_{I\mu} J_I^\mu
+\delta\phi_I^{\dot A}{}_{\dot B} K_I^{\dot B}{}_{\dot A}
\nonumber\\&&
+\delta F_{I+1}^A{}_B\wt\mu_I^B{}_A
+i\delta\lambda_{I+1A\dot B}\wt j_I^{A\dot B}
-\delta v_{I+1\mu}\wt J_I^\mu
-\delta\phi_{I+1}^{\dot A}{}_{\dot B}\wt K_I^{\dot B}{}_{\dot A},
\end{eqnarray}
where $S_{\rm hyper}^I$ is the
part of $S_{\rm hyper}$ including $(q_I,\psi_I)$.

$\mu$, $\wt\mu$, $j$, and $\wt j$ have been already given
in (\ref{mudef}) and (\ref{jdef}).
The other components are
\begin{eqnarray}
J_I^\mu
&=&iq_I^AD_\mu \ol q_{IA}-iD_\mu q_I^A \ol q_{IA}
+(\psi_{I\dot A}\gamma_\mu\ol\psi_I^{\dot A})
,\\
\wt J_I^\mu
&=&
-i\ol q_{IA}D^\mu q_I^A
+iD^\mu\ol q_{IA}q^A_I
-(\ol\psi_I^{\dot A}\gamma^\mu\psi_{I\dot A})
,\\
K_I^{\dot A}{}_{\dot B}
&=&
i\psi_{I\dot B}\ol\psi_I^{\dot A}
-\frac{i}{2}\delta^{\dot A}_{\dot B}\psi_{I\dot C}\ol\psi_I^{\dot C}
-\frac{1}{2}\nu_I^C{}_C \phi_I^{\dot A}{}_{\dot B}
-\frac{1}{2}\phi_I^{\dot A}{}_{\dot B}\nu_I^C{}_C
+q_I^C \phi_{I+1}^{\dot A}{}_{\dot B}\ol q_{IC}
,\\
\wt K_I^{\dot A}{}_{\dot B}
&=&
+i\ol\psi_I^{\dot A}\psi_{I\dot B}
-\frac{i}{2}\delta^{\dot A}_{\dot B}\ol\psi_I^{\dot C}\psi_{I\dot C}
+\frac{1}{2}\wt\nu_I^C{}_C\phi_{I+1}^{\dot A}{}_{\dot B}
+\frac{1}{2}\phi_{I+1}^{\dot A}{}_{\dot B}\wt\nu_I^C{}_C
-\ol q_{IC}\phi_I^{\dot A}{}_{\dot B}q_I^C.
\end{eqnarray}
The ${\cal N}=4_{\rm YM}$ supersymmetry transformation
of
$\mu$, $\wt\mu$, $j$, and $\wt j$ are
\begin{eqnarray}
\delta\mu_I^A{}_B
&=&
i\xi_{B\dot C}j_I^{A\dot C}
-\frac{i}{2}\delta^A_B\xi_{D\dot C}j_I^{D\dot C},\\
\delta\wt\mu_I^A{}_B
&=&
i\xi_{B\dot C}\wt j_I^{A\dot C}
-\frac{i}{2}\delta^A_B\xi_{D\dot C}\wt j_I^{D\dot C},\\
\delta j_I^{A\dot B}
&=&
-i\gamma_\mu\xi^{A\dot B}J_I^\mu
+2\gamma_\mu\xi^{C\dot B} D^\mu \mu_I^A{}_C
-2\xi^{A\dot C}K_I^{\dot B}{}_{\dot C}
+2\xi^{C\dot D}[\mu_I^A{}_C,\phi_I^{\dot B}{}_{\dot D}],\\
\delta\wt j_I^{A\dot B}
&=&
-i\gamma^\mu\xi^{A\dot B}\wt J_{I\mu}
+2\gamma^\mu\xi^{C\dot B} D_\mu\wt\mu_I^A{}_C
-2\xi^{A\dot C}\wt K_I^{\dot B}{}_{\dot C}
+2\xi^{C\dot D}[\wt\mu_I^A{}_C,\phi_{I+1}^{\dot B}{}_{\dot D}],\\
\delta J_I^\mu
&=&
\xi_{A\dot B}\gamma^{\mu\nu}D_\nu j_I^{A\dot B}
-\sqrt2\xi_{A\dot B}\gamma^\mu q_I^A\ol\Psi_I^{\dot B}
+\sqrt2\xi^{A\dot B}\gamma^\mu\Psi_{I\dot B}\ol q_{IA}
\nonumber\\&&
-[\xi_{B\dot A}\gamma^\mu j_I^{B\dot C},\phi_I^{\dot A}{}_{\dot C}]
+2[\xi_{C\dot B}\gamma^\mu\lambda_I^{A\dot B},\mu_I^C{}_A]
,\\
\delta\wt J_I^\mu
&=&
\xi_{A\dot B}\gamma^{\mu\nu}D_\nu\wt j_I^{A\dot B}
-\sqrt2\xi_{A\dot B}\gamma^\mu\ol\Psi_I^{\dot B}q_I^A
+\sqrt2\xi^{A\dot B}\gamma^\mu\ol q_{IA}\Psi_{I\dot B}
\nonumber\\&&
-[\xi_{C\dot B}\gamma^\mu\wt j_I^{C\dot A},\phi_{I+1}^{\dot B}{}_{\dot A}]
+2[\xi_{A\dot B}\gamma^\mu\lambda_{I+1}^{C\dot B},\wt\mu_I^A{}_C]
,\\
\delta K_I^{\dot A}{}_{\dot B}
&=&
-i\xi_{C\dot B}\gamma^\mu D_\mu j_I^{C\dot A}
+\sqrt2i\xi_{C\dot B}q_I^C\ol\Psi_I^{\dot A}
+\sqrt2i\xi^{C\dot A}\Psi_{I\dot B}\ol q_{IC}
\nonumber\\&&
-i[\xi_{D\dot C}j_I^{D\dot A}-\tr,\phi_I^{\dot C}{}_{\dot B}]
-2i[\xi_{D\dot B}\lambda_I^{C\dot A},\mu_I^D{}_C]
-\tr,\\
\delta\wt K_I^{\dot A}{}_{\dot B}
&=&
-i\xi_{C\dot B}\gamma^\mu D_\mu\wt j_I^{C\dot A}
+\sqrt2i\xi_{C\dot B}\ol\Psi_I^{\dot A}q_I^C
+\sqrt2i\xi^{C\dot A}\ol q_{IC}\Psi_{I\dot B}
\nonumber\\&&
-i[\xi_{D\dot C}\wt j_I^{D\dot A}-\tr,\phi_{I+1}^{\dot C}{}_{\dot B}]
-2i[\xi_{C\dot B}\lambda_{I+1}^{D\dot A},\wt\mu_I^C{}_D]-\tr.
\end{eqnarray}
These components are transformed among them linearly
up to the equation of motion of $\psi_{IA}$
given in (\ref{psieom}).



\end{document}